\documentclass[conference]{IEEEtran}
\IEEEoverridecommandlockouts
\usepackage{cite}
\usepackage{url}
\usepackage{hyperref}
\usepackage{float}
\usepackage{wrapfig}
\usepackage{amsmath,amssymb,amsfonts}
\usepackage{algorithmic}
\usepackage{graphicx}
\usepackage{textcomp}
\usepackage{xcolor}
\usepackage{interval}
\usepackage{dirtytalk}
\usepackage{subcaption}
\usepackage{paralist}

\def\BibTeX{{\rm B\kern-.05em{\sc i\kern-.025em b}\kern-.08em
    T\kern-.1667em\lower.7ex\hbox{E}\kern-.125emX}}
\begin{document}

\title{Redistribution contracts for transaction fees in blockchain protocols
}

\title{Fee-Redistribution Smart Contracts for Transaction-Fee-Based Regime of Blockchains with the Longest Chain Rule}

\author{\IEEEauthorblockN{Rastislav Budinsk\'{y}}
\IEEEauthorblockA{\textit{Brno University of Technology} \\
\textit{Faculty of Information Technology}\\
xbudin05@fit.vutbr.cz}
\and
\IEEEauthorblockN{Ivan Homoliak}
\IEEEauthorblockA{\textit{Brno University of Technology} \\
\textit{Faculty of Information Technology}\\
ihomoliak@fit.vutbr.cz}
\and
\IEEEauthorblockN{Ivana Stan\v{c}\'{i}kov\'{a}}
\IEEEauthorblockA{\textit{Brno University of Technology} \\
	\textit{Faculty of Information Technology}\\
	istancikova@fit.vutbr.cz}
\thanks{$\copyright$ 2023 IEEE.  Personal use of this material is permitted.  Permission from IEEE must be obtained for all other uses, in any current or future media, including reprinting/republishing this material for advertising or promotional purposes, creating new collective works, for resale or redistribution to servers or lists, or reuse of any copyrighted component of this work in other works.}
}

\maketitle

			\vspace{-1.9cm}
\begin{abstract}
	
In this paper, we review the undercutting attacks in the transaction-fee-based regime of proof-of-work (PoW) blockchains with the longest chain fork-choice rule.
Next, we focus on the problem of fluctuations in mining revenue and the mining gap -- i.e., a situation, in which the immediate reward from transaction fees does not cover miners' expenditures.

To mitigate these issues, we propose a solution that splits transaction fees from a mined block into two parts -- (1) an instant reward for the miner of a block and (2) a deposit sent to one or more fee-redistribution smart contracts ($\mathcal{FRSC}$s) that are part of the consensus protocol.
At the same time, these redistribution smart contracts reward the miner of a block with a certain fraction of the accumulated funds of the incoming fees over a predefined time.
This setting enables us to achieve several interesting properties that are beneficial for the incentive stability and security of the protocol.

With our solution, the fraction of \textsc{Default-Compliant} miners who strictly do not execute undercutting attack is lowered from the state-of-the-art result of 66\% to 30\%.
\end{abstract}

\section{Introduction}
\label{sec:introduction}
Cryptocurrencies provide block rewards to incentivize miners in producing new blocks.
Transaction fees also contribute to the revenue of miners, motivating them to include transactions in blocks.
Miners maximize their profits by prioritizing the transactions with the highest fees.
In Bitcoin and its numerous clones~\cite{hum2020coinwatch}, the block reward is divided by two approx. every four years (i.e., after every 210k blocks), which will eventually result in a pure transaction-fee-based regime.

Before 2016, there was a belief that the dominant source of the miners' income does not impact the security of the blockchain.
However, Carlsten et al.~\cite{carlsten2016instability} pointed out the effects of the high variance of the miners' revenue per block caused by exponentially distributed block arrival time in transaction-fee-based protocols.
The authors showed that \emph{undercutting} (i.e., forking) a wealthy block is a profitable strategy for a malicious miner. 
Nevertheless, Daian et al.~\cite{daian2020flash} showed that this attack is viable even in blockchains containing traditional block rewards due to front-running competition of arbitrage bots who are willing to extremely increase transaction fees to earn Maximum Extractable Value profits.

In this paper, we focus on mitigation of the undercutting attack in transaction-fee-based regime of PoW blockchains.
We also discuss related problems present (not only) in transaction-fee-based regime.
In particular, we focus on minimizing the mining gap~\cite{carlsten2016instability,tsabary2018gap}, (i.e., the situation, where the immediate reward from transaction fees does not cover miners' expenditures) as well as balancing significant fluctuations in miners' revenue.

To mitigate these issues, we propose a solution that splits  transaction fees from a mined block into two parts -- (1) an instant reward for the miner and (2) a deposit sent into one or more fee-redistribution smart contracts ($\mathcal{FRSC}$s).
At the same time, these $\mathcal{FRSC}$s reward the miner of a block with a certain fraction of the accumulated funds over a fixed period of time (i.e., the fixed number of blocks).
This setting enables us to achieve several interesting properties that are beneficial for the stability and security of the protocol.

\subsubsection*{\textbf{Contributions}}
In detail, our contributions are as follows:
\begin{enumerate}
	\item We propose an approach that normalizes the mining rewards coming from transaction fees by one or more $\mathcal{FRSC}$s that perform moving average on a certain portion of the transaction fees.
	
	\item We evaluate our approach using various fractions of the transaction fees from a block distributed between a miner and $\mathcal{FRSC}$s.
	We experiment with the various numbers and lengths of $\mathcal{FRSC}$s, and we demonstrate that usage of multiple $\mathcal{FRSC}$s of various lengths has the best advantages mitigating the problems we are addressing; however, even using a single $\mathcal{FRSC}$ is beneficial.

	\item We demonstrated that with our approach, the mining gap can be minimized since the miners at the beginning of the mining round can get the reward from $\mathcal{FRSC}$s, which stabilizes their income.
	
	\item We empirically demonstrate that using our approach the threshold of \textsc{Default-Compliant} miners who strictly do not execute undercutting attack is lowered from 66\% (as reported in the original work~\cite{carlsten2016instability}) to 30\%. 
	
\end{enumerate}

\section{Problem Definition}
\label{sec:problemdefinition}
In transaction fee-based regime schemes, a few problems have emerged, which we can observe even nowadays in Bitcoin protocol~\cite{carlsten2016instability}.
We have selected three main problems and aim to lower their impact for protocols relying only on transaction fees. 
In detail, we focus on the following problems:

\begin{enumerate}
	\item \textbf{Undercutting attack.}
	In this attack, a malicious miner attempts to obtain transaction fees by re-mining a top block of the longest chain, and thus motivates other miners to mine on top of her block~\cite{carlsten2016instability}.
	In detail, consider a situation, where an honest miner mines a block containing transaction with extremely high transaction fees. 
	The malicious miner can fork this block while he leaves some portion of the ``generous'' transactions un-mined in the mempool.
	These transactions motivate other miners to mine on top of the attacker's chain, and thus undercut the original block. 
	Such a malicious behavior might result in higher orphan rate, unreliability of the system, and double spending.
	
	\item \textbf{The mining gap.}
	As discussed in~\cite{carlsten2016instability}, the problem of mining gap arises once the mempool does not contain enough transaction fees to motivate miners in mining.
	Suppose a miner succeeds at mining a new block shortly after the previous block was created, which can happen due to well known exponential distribution of block creation time in PoW blockchains.
	Therefore, the miner might not receive enough rewards to cover his expenses because most of the transactions from the mempool were included in the previous block, while new transactions might not have yet arrived or have small fees.
	Consequently, the miners are motivated to postpone mining until the mempool is reasonably filled with enough transactions (and their fees).
	The mining gap was also analyzed by the simulation approach in the work of Tsabary and Eyal~\cite{tsabary2018gap}, who further demonstrated that this mining gap incentivizes larger mining coalitions, which impacts decentralization.
	
	\item \textbf{Varying transaction fees over time.}
	In the transaction-fee-based regime, any fluctuation in transaction fees directly affects the miners' revenue.
	High fluctuation of transaction fees during certain time frames, e.g., in a span of a day or a week~\cite{b5}, can lead to an undesirable lack of predictability in rewards of miners and indirectly affect the security of the underlying protocol.

\end{enumerate}

\section{Proposed approach}
\label{sec:proposedapproach}

In this section, we describe our proposed solution in more detail.
\begin{figure}[t]
	\includegraphics[width=0.45\textwidth]{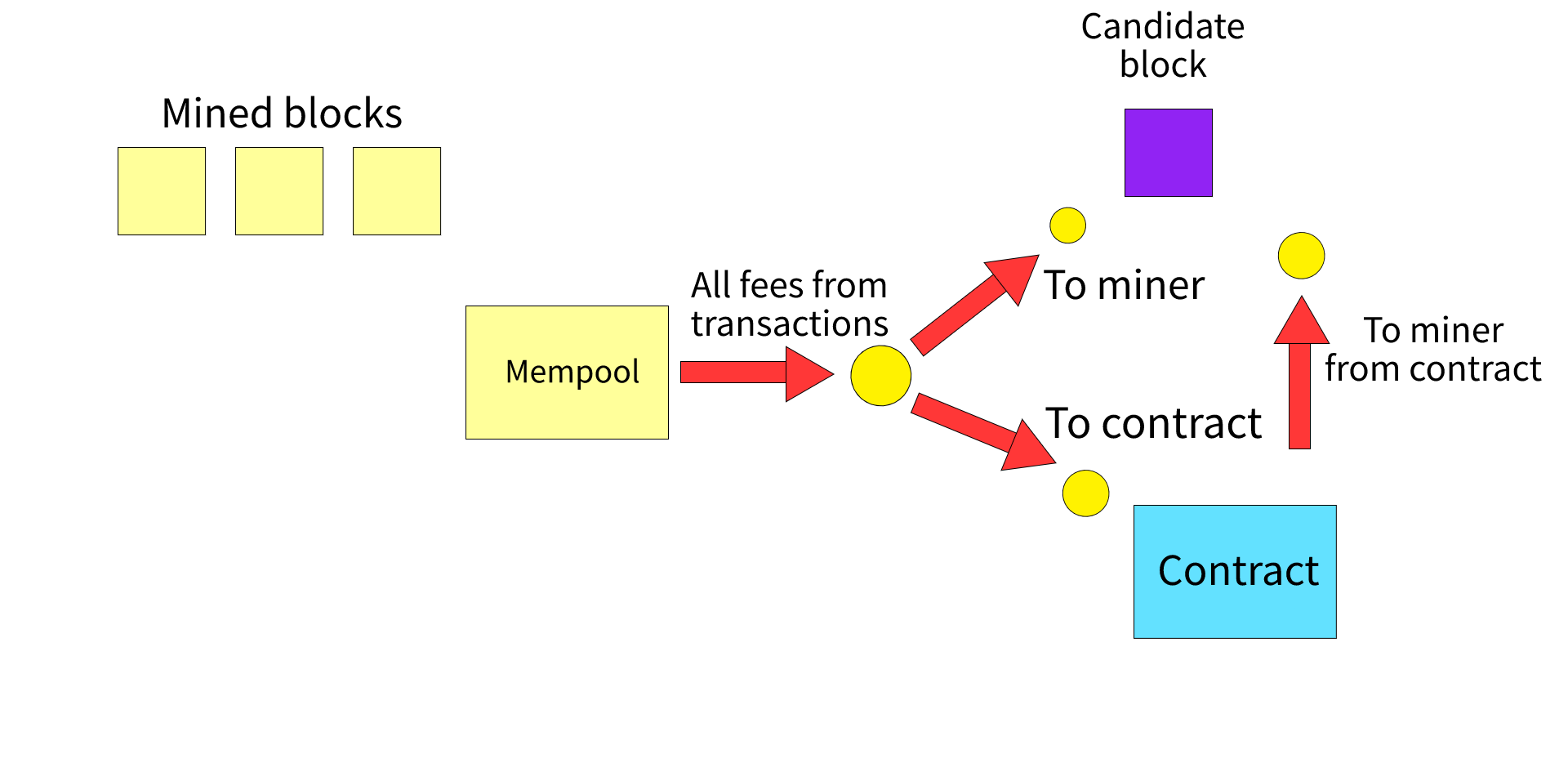}
			\vspace{-0.7cm}
	\caption{Overview of our solution. }
	\label{fig:overview}
\end{figure}
Our proposed solution collects a percentage of transaction fees in a native cryptocurrency from the mined blocks into one or multiple fee-redistribution smart contracts (i.e., $\mathcal{FRSC}$s).
Miners of the blocks who must contribute to these contracts are at the same time rewarded from them, while the received reward approximates a moving average of the incoming transaction fees across the fixed sliding window of the blocks.
The fraction of transaction fees (i.e., $\mathbb{C}$) from the mined block is sent to $\mathcal{FRSC}$ and the remaining fraction of transaction fees (i.e., $\mathbb{M}$) is directly assigned to the miner, such that $ \mathbb{C} + \mathbb{M} = 1.$

The role of $\mathbb{M}$ is to incentivize the miners in prioritization of the transactions with the higher fees while the role of $\mathbb{C}$ is to mitigate the problems of undercutting attacks and the mining gap.

\medskip\noindent
\subsubsection*{\textbf{Overview}}
We depict the overview of our approach in \autoref{fig:overview}, and it consists of the following steps:
\begin{enumerate}
	
	\item Using $\mathcal{FRSC}$, the miner calculates the reward for the next block $B$ (i.e., $nextClaim(\mathcal{FRSC})$ -- see \autoref{eq:nextClaim}) that will be payed by $\mathcal{FRSC}$ to the miner of that block.
	
	\item The miner mines the block $B$ using the selected set of the highest fee transactions from her mempool.
	
	\item The mined block $B$ directly awards a certain fraction of the transaction fees (i.e., $B.fees ~*~ \mathbb{M}$) to the miner and the remaining part (i.e., $B.fees ~*~ \mathbb{C}$) to $\mathcal{FRSC}$.
	
	\item The miner obtains $nextClaim$ from $\mathcal{FRSC}$.
	
\end{enumerate}
Our approach is embedded into the consensus protocol, and therefore consensus nodes are obliged to respect it in order to ensure that their blocks are valid.
It can be implemented with standard smart contracts of the blockchain platform or within the native code of the consensus protocol.

\subsection{Prioritization of High-Fee Transactions}
In the environment with constant transaction fees, a miner would receive the same amount with or without our solution.
However, in public blockchains (especially with transaction-fee based regime) there exist a mechanism to ensure prioritizaiton in processing of transactions with higher fees, which might result into fluctuations in rewards of the miners.
In our approach, we preserve the transaction prioritization since we directly attribute a part of the transaction fees to the miner (i.e., $\mathbb{M}$).

\subsection{Fee-Redistribution Smart Contracts}\label{sub:redistrib}
We define the fee-redistribution smart contract as 
\begin{eqnarray}
\mathcal{FRSC} = (\nu, \lambda, \rho), 
\end{eqnarray}
where
$\nu$ is the accumulated amount of tokens in the contract, $\lambda$ denotes the size of $\mathcal{FRSC}$' sliding window in terms of the number of preceding blocks that contributed to $\nu$,
and $\rho$ is the parameter defining the ratio for redistribution of incoming transaction fees among multiple contracts, while the sum of $\rho$ across all $\mathcal{FRSC}$s must be equal to 1:
\begin{eqnarray}\label{eq:frsc-redistrib-ratios}
	\sum_{x ~\in~ \mathcal{FRSC}s} x.\rho &=& 1.
\end{eqnarray}
In contrast to a single $\mathcal{FRSC}$, multiple $\mathcal{FRSC}$s enable better adjustment of compensation to miners during periods of higher transaction fee fluctuations or in an unpredictable environment (we show this in \autoref{sec:exp3}).

\medskip
We denote the state of $\mathcal{FRSC}$s at the blockchain height $H$ as $\mathcal{FRSC}_{[H]}$.
Then, we determine the reward from $\mathcal{FRSC}_{[H]} \in \mathcal{FRSC}s_{[H]}$ for the miner of the next block with height $H+1$ as follows:
\begin{equation}
	\partial Claim_{[H+1]}^{\mathcal{FRSC}_{[H]}} = \frac{\mathcal{FRSC}_{[H]}.\nu}{\mathcal{FRSC}_{[H]}.\lambda},
\end{equation}
while the reward obtained from all $\mathcal{FRSC}$s is
\begin{equation} \label{eq:nextClaim}	
nextClaim_{[H+1]} = \sum_{\mathcal{X}_{[H]} ~\in~ \mathcal{FRSC}s_{[H]}^{}} \partial Claim_{[H+1]}^{\mathcal{X}_{[H]}}.
\end{equation}

\noindent
Then, the total reward of the miner who mined the block $B_{[H+1]}$ with all transaction fees $B_{[H+1]}.fees$ is
\begin{equation} \label{eq:rewardtotal}
	rewardT_{[H+1]} = nextClaim_{[H+1]} + \mathbb{M} * B_{[H+1]}.fees.
\end{equation}
The new state of contracts at the height $H + 1$ is 
\begin{eqnarray}
	\mathcal{FRSC}s_{[H+1]} = \{\mathcal{X}_{[H+1]}(\nu, \lambda, \rho)\ ~|~
\end{eqnarray}
\begin{eqnarray}
	\lambda &=& \mathcal{X}_{[H]}.\lambda,\\
	\rho &=& \mathcal{X}_{[H]}.\rho,\\
	\nu &=& \mathcal{X}_{[H]}.\nu - \partial Claim_{[H+1]} +  deposit * \rho,\\
	deposit &=& B_{[H+1]}.fees * \mathbb{C}\},
\end{eqnarray}
where $deposit$ represents the fraction $\mathbb{C}$ of all transaction fees from the block $B_{[H+1]}$ that are deposited across all $\mathcal{FRSC}$s in ratios respecting \autoref{eq:frsc-redistrib-ratios}.

\subsection{Example}
We present an example using Bitcoin~\cite{nakamoto2008bitcoin} to demonstrate our approach.
We assume that the current height of the blockchain is $H$, and we utilize only a single $\mathcal{FRSC}$ with the following parameters:
\[
\mathcal{FRSC}_{[H]} = (2016, 2016, 1).
\]
We set $\mathbb{M} = 0.4$ and $\mathbb{C} = 0.6$, which means a miner directly obtains 40\% of the $B_{[H+1]}.fees$ and $\mathcal{FRSC}$ obtains remaining 60\%.

\medskip
\noindent
Next, we compute the reward from $\mathcal{FRSC}$ obtained by the miner of the block with height $H+1$ as 
\begin{equation*}
	\partial Claim_{[H+1]} = \frac{\mathcal{FRSC}_{[H]}.\nu}{\mathcal{FRSC}_{[H]}.\lambda} = \frac{2016}{2016} = 1~\text{BTC},
\end{equation*}
resulting into 
\begin{equation*}
nextClaim_{[H+1]} = \partial Claim_{[H+1]}~=~1~\text{BTC}.
\end{equation*}

\noindent
Further, we assume that the total reward collected from transactions in the block with height $H+1$ is $B_{[H+1]}.fees = 2$ BTC.
Hence, the total reward obtained by the miner of the block $B_{[H+1]}$ is
\begin{eqnarray*}
	rewardT_{[H+1]} &=& nextClaim_{[H+1]}  + \mathbb{M} * B_{[H+1]}.fees \\
	&=& 1 + 0.4 * 2 \\
	&=& 1.8 ~\text{BTC},
\end{eqnarray*}
and the contribution of transaction fees from $B_{[H+1]}$ to the $\mathcal{FRSC}$ is 
\[
deposit = B_{[H+1]}.fees * \mathbb{C} = 1.2 ~\text{BTC}.
\]
Therefore, the value of $\nu$ in $\mathcal{FRSC}$ is updated at height H~+~1 as follows:
\begin{eqnarray*}
	v_{[H+1]} &=& \mathcal{FRSC}_{[H]}.\nu - nextClaim_{[H+1]} + deposit \\
	&=& 2016 - 1 + 1.2  ~\text{BTC} \\
	&=& 2016.2 ~\text{BTC}.
\end{eqnarray*}

\noindent
\subsubsection*{\textbf{Traditional Way in Tx-Fee Regime}}
In traditional systems (running in transaction-fee regime) $rewardT_{[H+1]}$ would be equal to the sum of all transaction fees $B_{[H+1]}.fees$ (i.e., $2$ BTC); hence, using $\mathbb{M} = 1$. 
In our approach, $rewardT_{[H+1]}$ can only be equal to the sum of all transaction fees in the block $B_{[H+1]}$, if:
\begin{eqnarray}
	B_{[H+1]}.fees = \frac{nextClaim_{[H+1]}}{\mathbb{C}}.
\end{eqnarray}
In our example, a miner can mine the block $B_{[H+1]}$ while obtaining the same total reward as the sum of all transaction fees in the block if the transactions carry 1.66 BTC in fees:
\begin{equation*}
B_{[H+1]}.fees = \frac{1}{0.6} = 1.66 ~\text{BTC}.
\end{equation*}

\subsection{Initial Setup of $\mathcal{FRSC}$s Contracts}
To enable an even start, we propose to initiate $\mathcal{FRSC}$s of our approach by a genesis value.
The following formula calculates the genesis values per $\mathcal{FRSC}$ and initializes starting state of $\mathcal{FRSC}s_{[0]}$:
\begin{equation} \label{eq:setup}
	\{\mathcal{FRSC}_{[0]}^{x}(\nu, \lambda, \rho)\ |\ \nu = \overline{fees} * \mathbb{C} * \rho * \lambda\},
\end{equation}
where $\overline{fees}$ is the expected average of incoming fees. 

\section{Implementation}
\label{sec:implementation}
Our implementation is based on Bitcoin Mining Simulator~\cite{bitcoin_mining_simulator:kalodner}, introduced in~\cite{carlsten2016instability}, which was adjusted to our needs.

\subsubsection{Changes in Simulator}

We have created a configuration file to simulate custom scenarios of incoming transactions instead of the original design~\cite{carlsten2016instability} fees.
We added an option to switch simulation into a mode with a full mempool, and thus bound the total fees (and consequently the total number of transactions) that can be earned within a block -- 
this mostly relates to blocks whose mining takes longer time than the average time to mine a block.\footnote{Note that the original simulator~\cite{carlsten2016instability} assumes that the number of transactions (and thus the total fees) in the block is constrained only by the duration of a time required to mine the block, which was also criticized in~\cite{gong2022towards}.}

Next, we moved several parameters to arguments of the simulator with the intention to eliminate the need of recompilation the program frequently, and therefore simplify the process of running various experiments with the simulator.
Finally, we integrated our $\mathcal{FRCS}$-based solution into the simulator.
$\mathcal{FRSC}$s are initiated from a corresponding configuration file.
The source code of our modified simulator is available at anonymous link \url{https://www.dropbox.com/s/gwlbedq47csthqp/sources.zip?dl=0}.

\section{Evaluation}
\label{sec:evaluation}
We evaluated our proof-of-concept implementation of $\mathcal{FRCS}$s on a long term scenario that we designed to demonstrate significant changes in the total transaction fees in the mempool evolving across the time.
This scenario is depicted in the resulting graphs of most of our experiments, represented by the ``\textit{Fees in mempool}'' series -- see \autoref{sec:exp1} and \autoref{sec:exp2}.

We experimented with different parameters and investigated how they influenced the total rewards of miners coming from $\mathcal{FRSC}$s versus the baseline without our solution.
Mainly these included a setting of $\mathbb{C}$ as well as different lengths $\lambda$ of $\mathcal{FRSC}$s.
Note that we used the value of transaction fees per block equal to 50 BTC, alike in the original paper introducing undercutting attacks~\cite{carlsten2016instability}.
Also, in all our experiments but the last one (i.e., \autoref{sec:exp4}), we enabled the full mempool option to ensure more realistic conditions.

\begin{figure*}[t]
	\vspace{-0.7cm}
	\centering
	\begin{subfigure}{0.40\textwidth}
		\includegraphics[width=\textwidth]{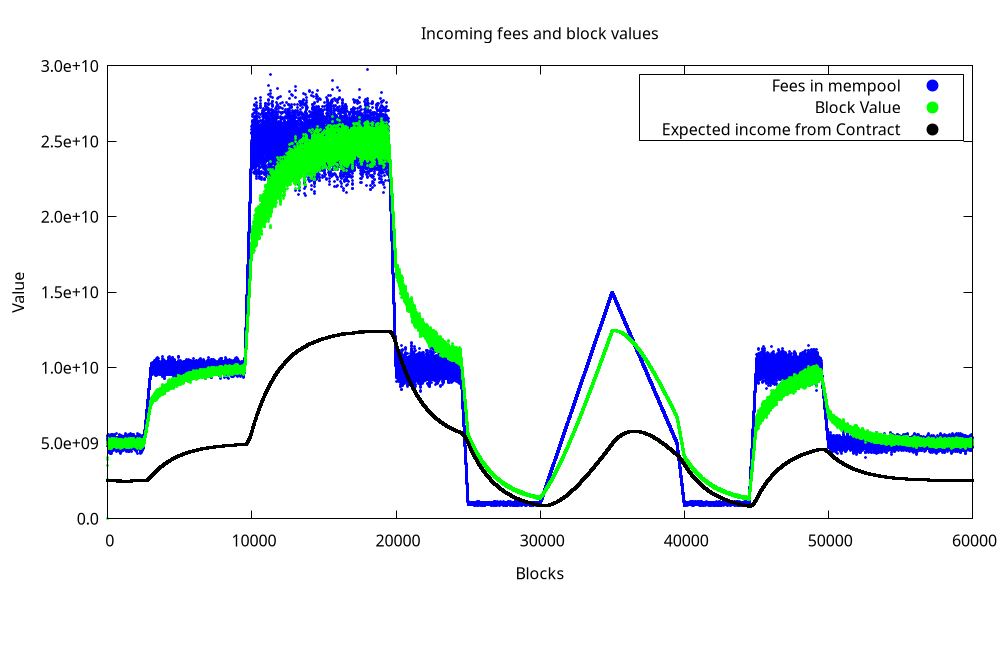}
			\vspace{-0.9cm}
		\caption{$\mathcal{FRSC}^{1}$ and $\mathbb{C} = 0.5$.}
	\end{subfigure}
	\begin{subfigure}{0.40\textwidth}
		\includegraphics[width=\textwidth]{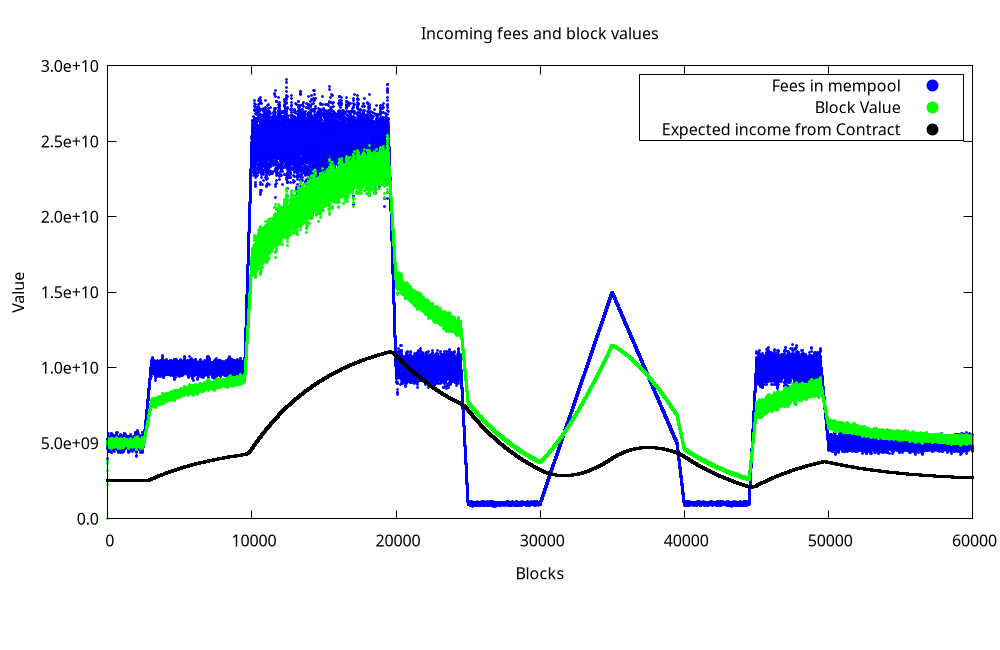}
					\vspace{-0.9cm}
		\caption{$\mathcal{FRSC}^{2}$ and $\mathbb{C} = 0.5$.}
	\end{subfigure}
	\begin{subfigure}{0.40\textwidth}
		\includegraphics[width=\textwidth]{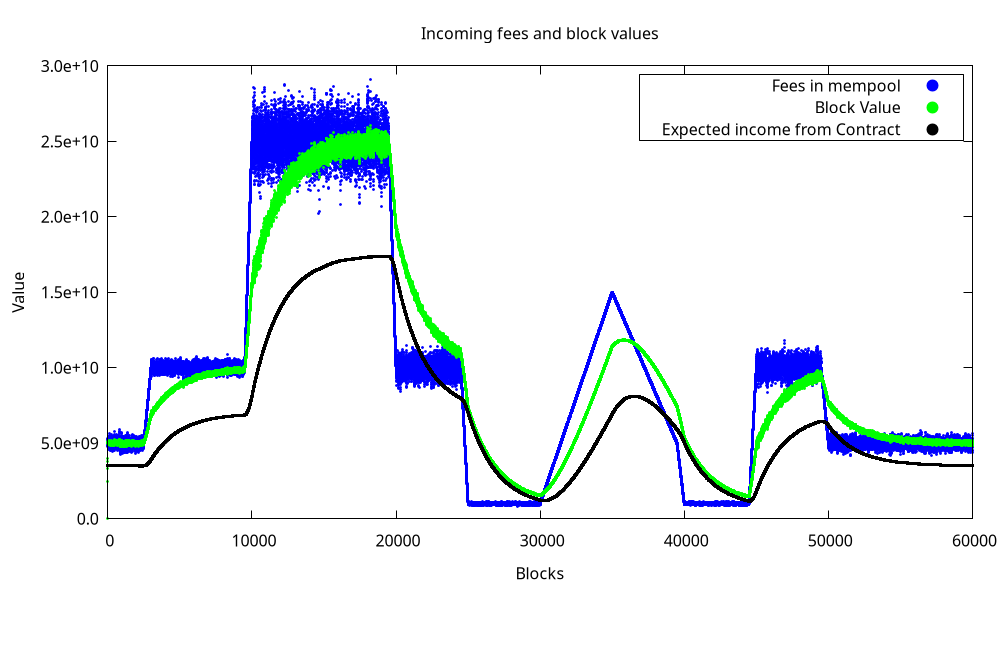}
					\vspace{-0.9cm}
		\caption{$\mathcal{FRSC}^{1}$ and $\mathbb{C} = 0.7$.}
	\end{subfigure}
	\begin{subfigure}{0.40\textwidth}
		\includegraphics[width=\textwidth]{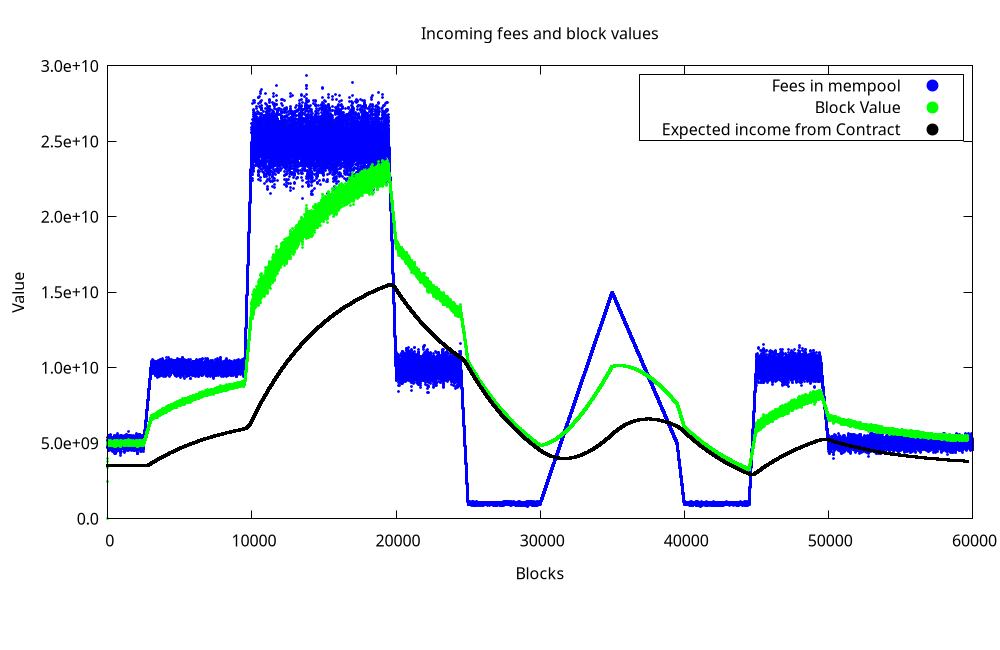}
					\vspace{-0.9cm}
		\caption{$\mathcal{FRSC}^{2}$ and $\mathbb{C} = 0.7$.}
	\end{subfigure}
	
	\begin{subfigure}{0.40\textwidth}
		\includegraphics[width=\textwidth]{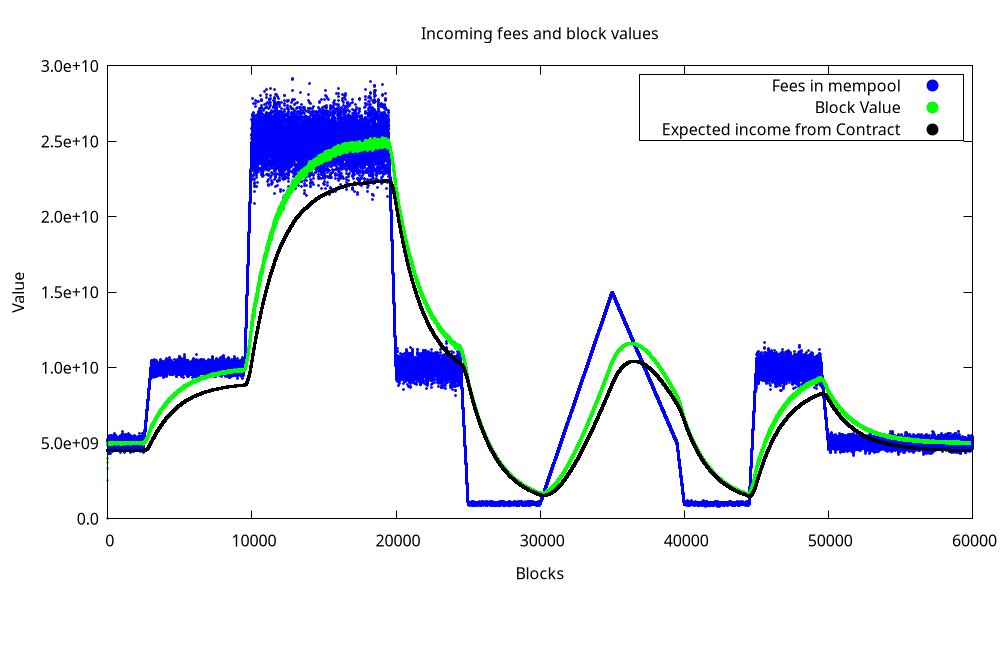}
					\vspace{-0.9cm}
		\caption{$\mathcal{FRSC}^{1}$ and $\mathbb{C} = 0.9$.}
	\end{subfigure}
	\begin{subfigure}{0.40\textwidth}
		\includegraphics[width=\textwidth]{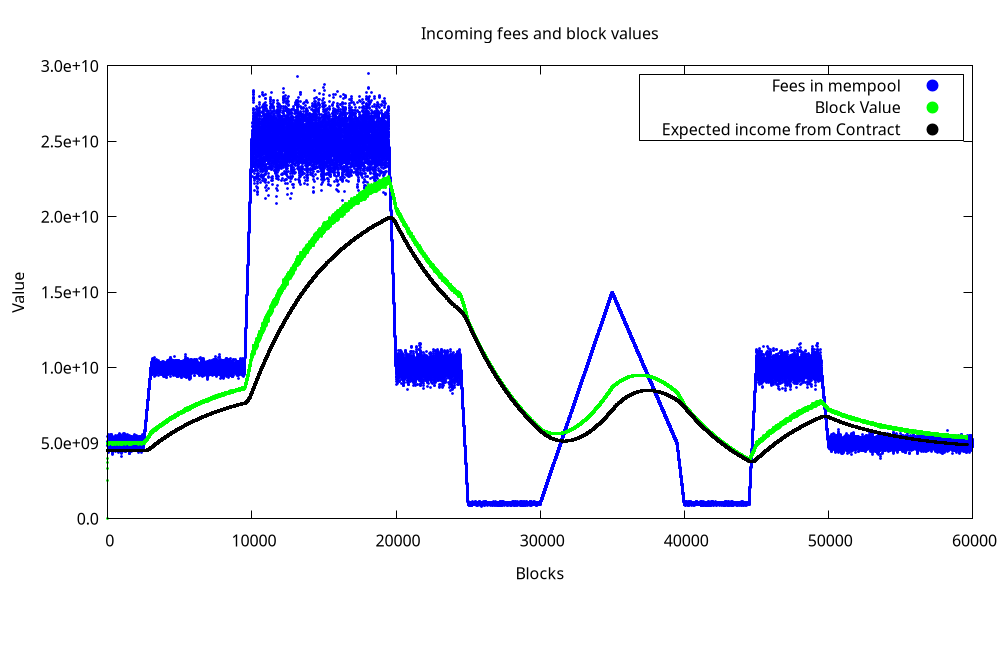}
					\vspace{-0.9cm}
		\caption{$\mathcal{FRSC}^{2}$ and $\mathbb{C} = 0.9$.}
					\vspace{-0.4cm}
	\end{subfigure}
	
	\caption{Experiment I investigating various $\mathbb{C}s$ and $\lambda$s of a single $\mathcal{FRSC}$, where
		$\mathcal{FRSC}^{1}.\lambda = 2016$ and 		$\mathcal{FRSC}^{2}.\lambda = 5600$.
		\textit{Fees in mempool} show the total value of fees in the mined block (i.e., representing the baseline).
		\textit{Block Value} is the reward a miner received in block $B$ as a sum of the fees he obtained directly (i.e. $\mathbb{M} * B.fees$) and the reward he got from $\mathcal{FRSC}$ (i.e., $nextClaim_{[H]}$).
		\textit{Expected income from Contract} represents the reward of a miner obtained from $\mathcal{FRSC}$ (i.e., $nextClaim_{[H]}$).}\label{fig:50tocontract}
\end{figure*}

\subsection{Experiment I}
\label{sec:exp1}

\subsubsection{\textbf{Methodology}}
The purpose of this experiment was to investigate the amount of the reward a miner received with our approach versus the baseline (i.e., the full reward is based on all transaction fees).
In this experiment, we investigated how $\mathbb{C}$ influences the total reward of the miner and how $\lambda$ of the sliding window averaged the rewards.
In detail, we created two independent $\mathcal{FRSC}$s with different $\lambda$ -- one was set to 2016 (i.e., $\mathcal{FRCS}^1$), and the second one was se to 5600 (i.e., $\mathcal{FRCS}^2$). 
We simulated these two $\mathcal{FRCS}$s with three different values of $\mathbb{C} \in \{0.5,~0.7,~0.9\}$.

\subsubsection{\textbf{Results}}
The results of this experiment are depicted in \autoref{fig:50tocontract}.
Across all runs of our experiment we can observe that $\mathcal{FRSC}^{2}$ adapts slower as compared to $\mathcal{FRSC}^{1}$, which leads to more significant averaging of the total reward paid to the miner.
Even though this is desired, we can see faster adaptation to changing environment by $\mathcal{FRSC}^{1}$, what a miner can expects from rising fees.

\begin{figure*}
	\vspace{-0.5cm}
	\centering
	\begin{subfigure}{0.32\textwidth}
		\includegraphics[width=\textwidth]{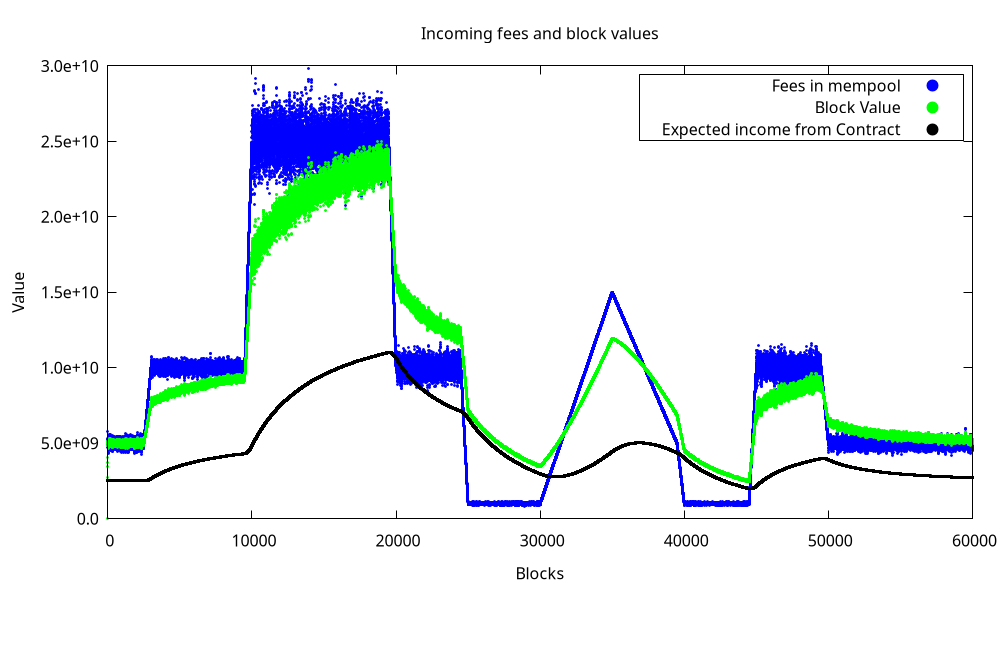}
		\caption{Scenario with 4 $\mathcal{FRSC}$s,\\ $\mathbb{C} = 0.5$.}
	\end{subfigure}
	\begin{subfigure}{0.32\textwidth}
		\includegraphics[width=\textwidth]{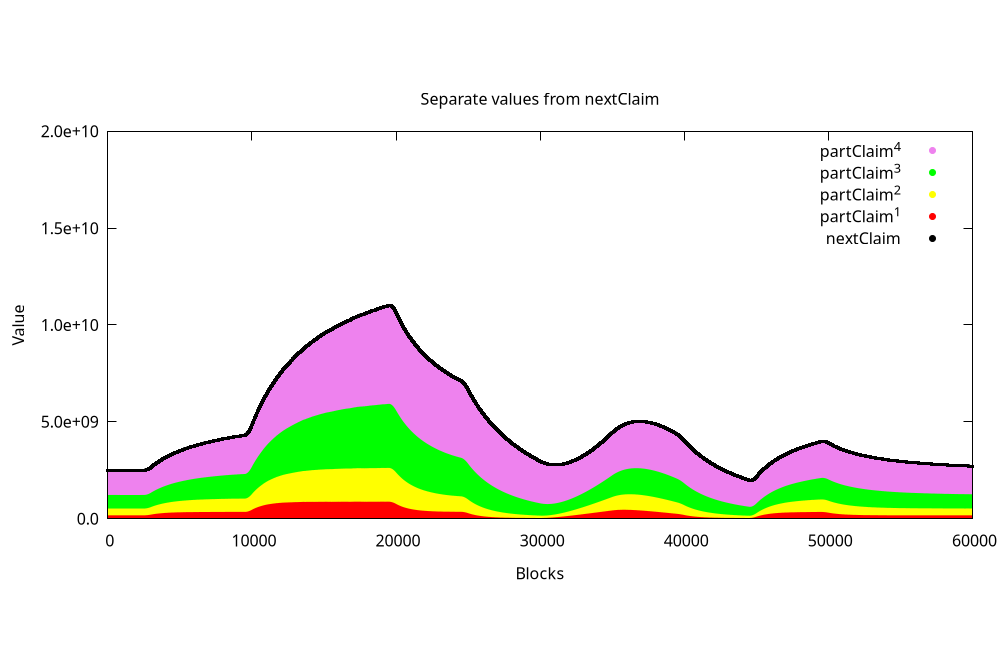}
		\caption{$\partial Claim$s and $nextClaim$, \\ $\mathbb{C} = 0.5$.}
	\end{subfigure}
	\begin{subfigure}{0.32\textwidth}
		\includegraphics[width=\textwidth]{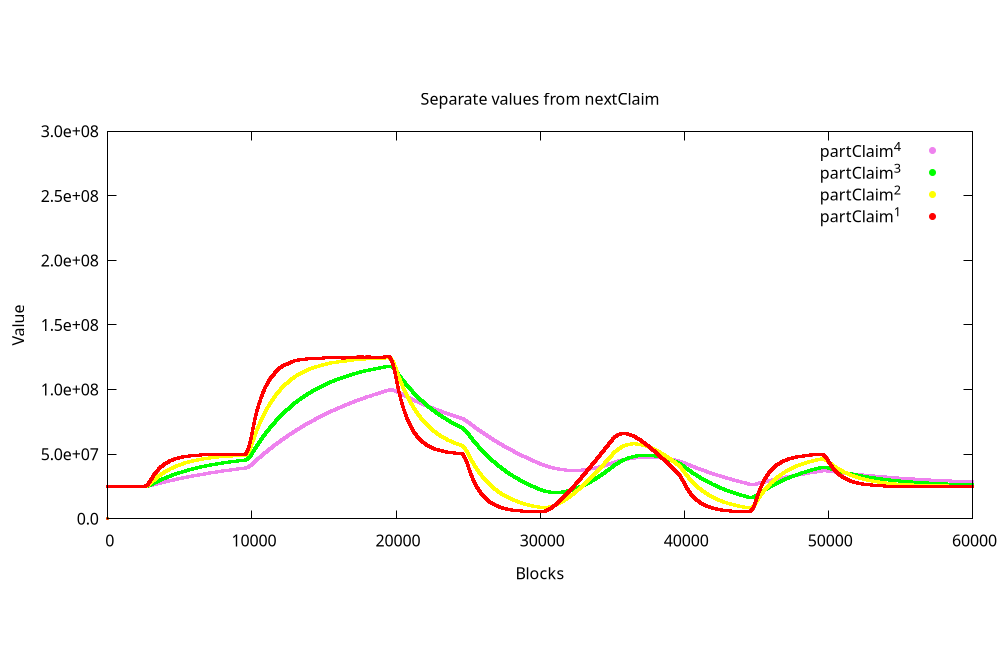}
		\caption{$\partial Claim$s normalized by $\rho$, \\ $\mathbb{C} = 0.5$.}
	\end{subfigure}

	\begin{subfigure}{0.32\textwidth}
		\includegraphics[width=\textwidth]{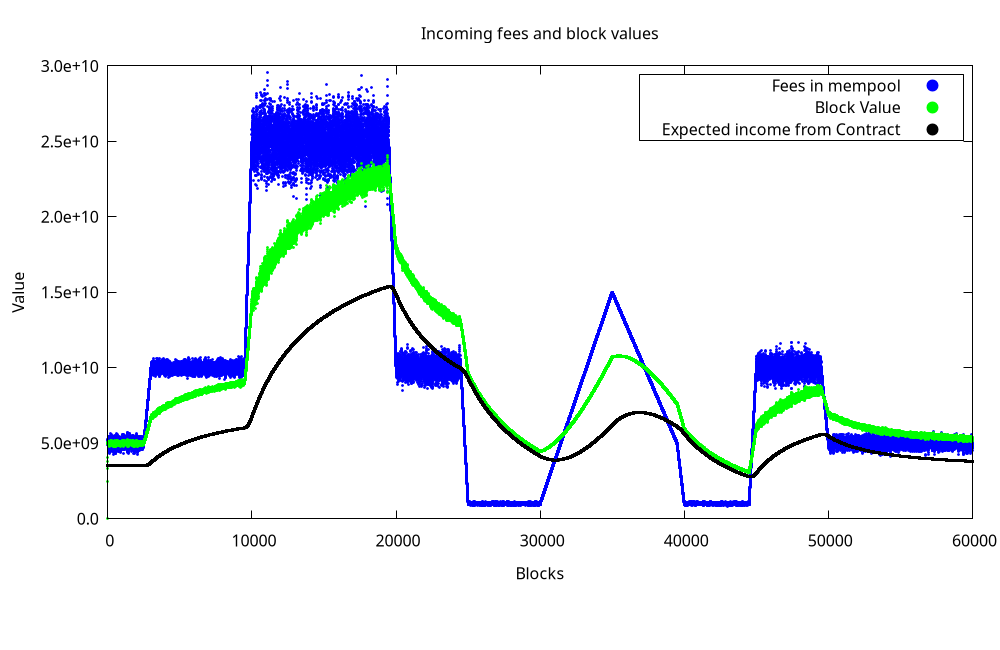}
		\caption{Scenario with 4 $\mathcal{FRSC}$s,\\ $\mathbb{C} = 0.7$.}
	\end{subfigure}
	\begin{subfigure}{0.32\textwidth}
		\includegraphics[width=\textwidth]{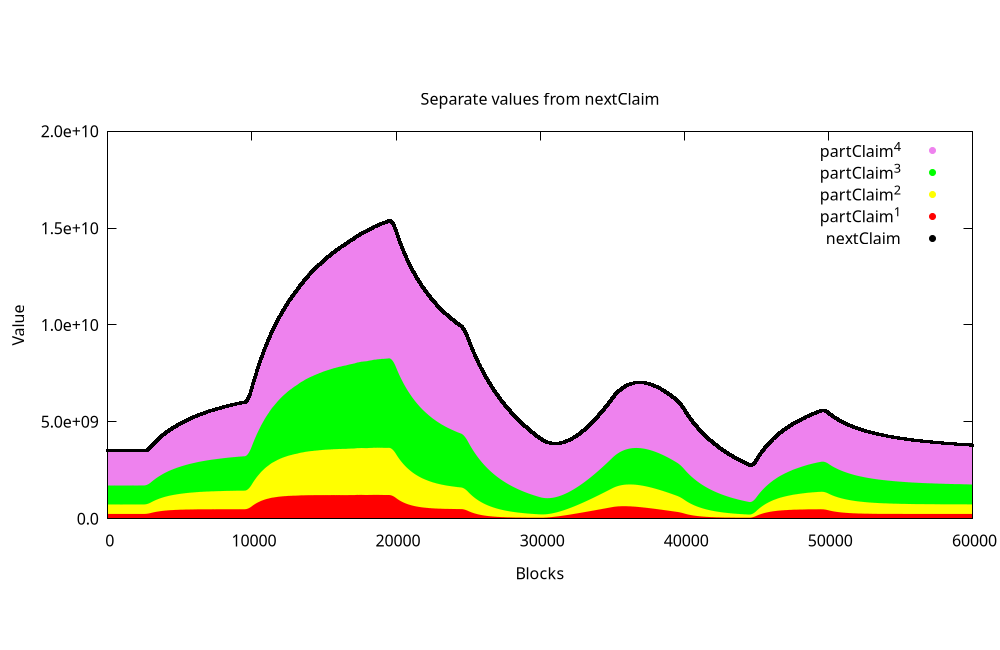}
		\caption{$\partial Claim$s and $nextClaim$, \\ $\mathbb{C} = 0.7$.}
	\end{subfigure}
	\begin{subfigure}{0.32\textwidth}
		\includegraphics[width=\textwidth]{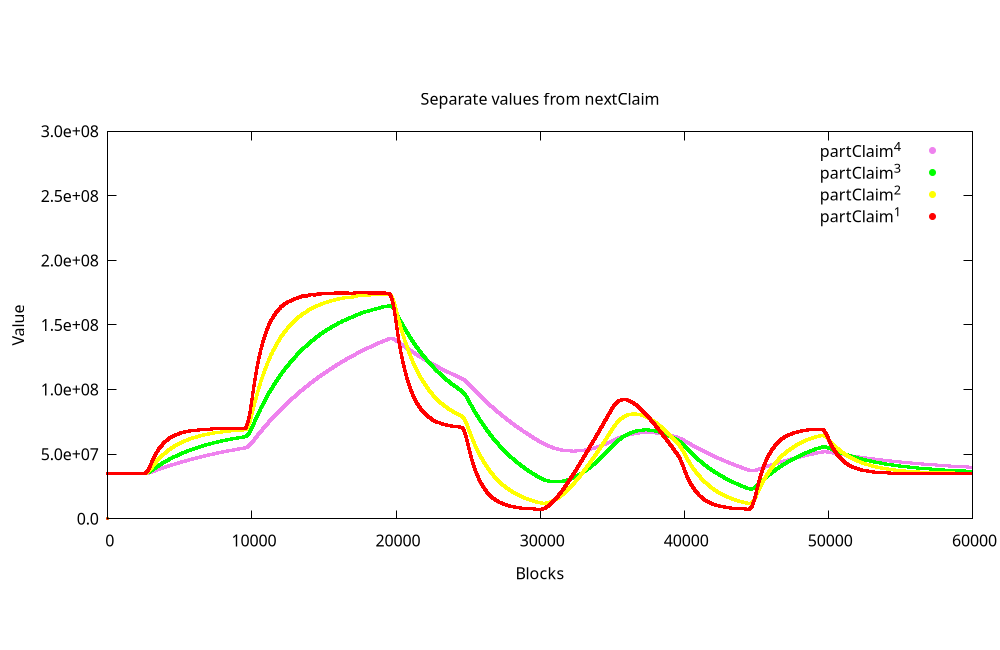}
		\caption{$\partial Claim$s normalized by $\rho$, \\ $\mathbb{C} = 0.7$.}
	\end{subfigure}

	\begin{subfigure}{0.32\textwidth}
		\includegraphics[width=\textwidth]{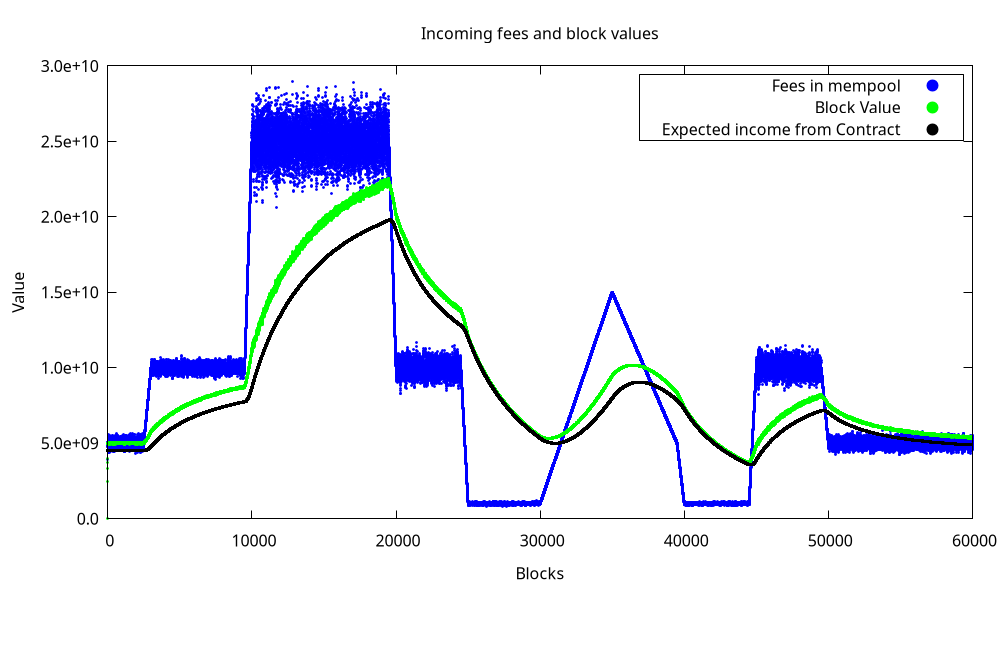}
		\caption{Scenario with 4 $\mathcal{FRSC}$s, \\ $\mathbb{C} = 0.9$.}
	\end{subfigure}
	\begin{subfigure}{0.32\textwidth}
		\includegraphics[width=\textwidth]{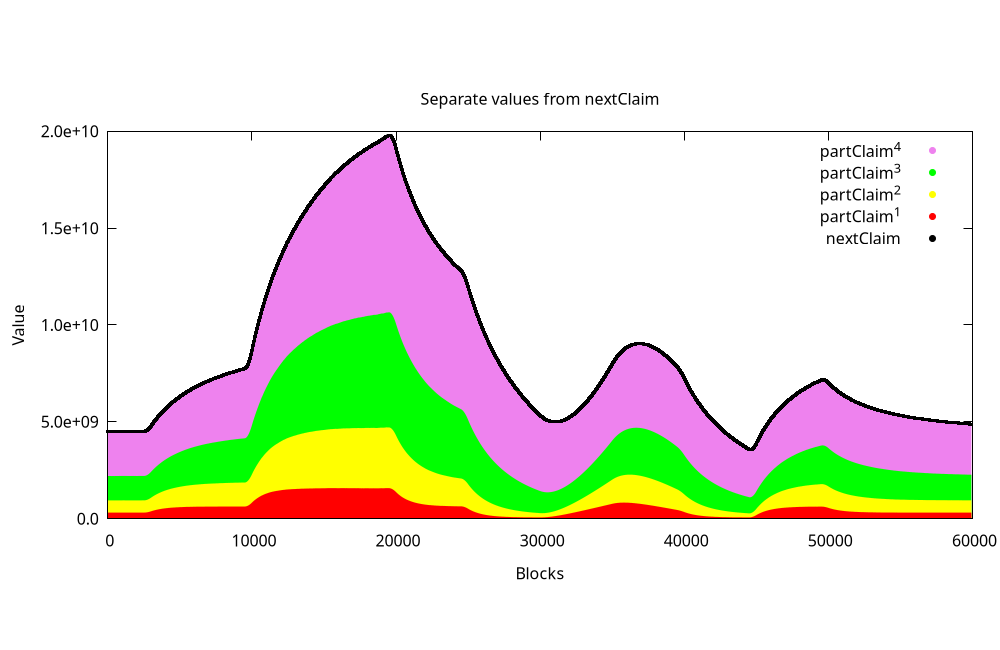}
		\caption{$\partial Claim$s and $nextClaim$, \\ $\mathbb{C} = 0.9$.}
	\end{subfigure}
	\begin{subfigure}{0.32\textwidth}
		\includegraphics[width=\textwidth]{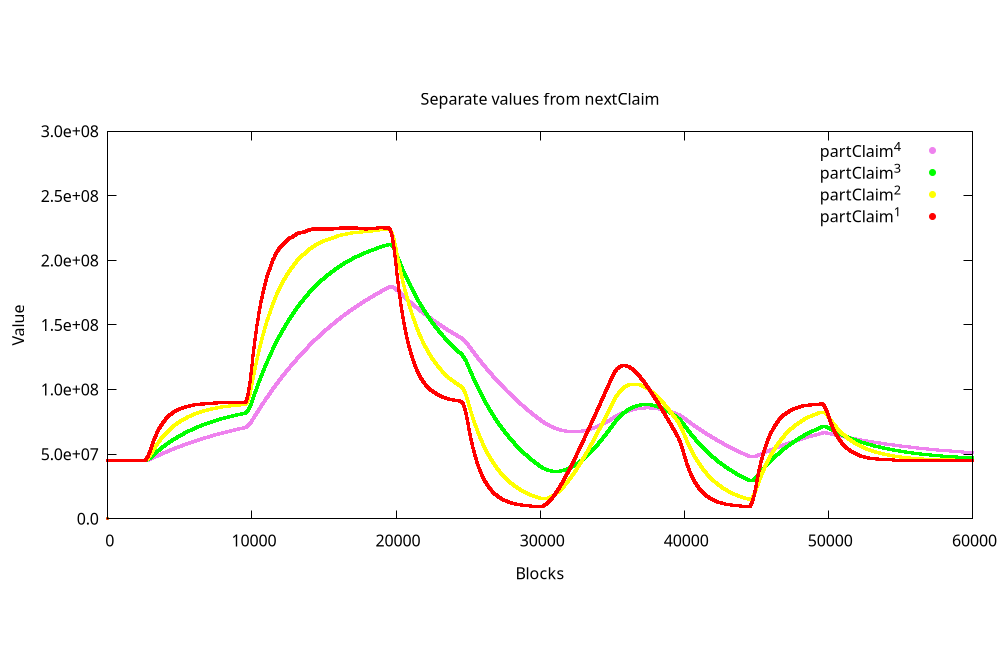}
		\caption{$\partial Claim$s normalized by $\rho$, \\ $\mathbb{C} = 0.9$.}
	\end{subfigure}
	
	\caption{Experiment II investigating various $\mathbb{C}$s in the setting with multiple $\mathcal{FRSC}$s with their corresponding $\lambda$ = $\{1008, 2016, 4032, 8064\}$ and $\rho$ = $\{0.07, 0.14, 0.28, 0.51\}$. $\partial Claim$s represents contributions of individual $\mathcal{FRSC}$s to the total reward of the miner (i.e., its $nextClaim$ component).}\label{fig:exp-II}
\end{figure*}

\begin{figure*}
	\centering
	\begin{subfigure}{0.32\textwidth}
		\includegraphics[width=\textwidth]{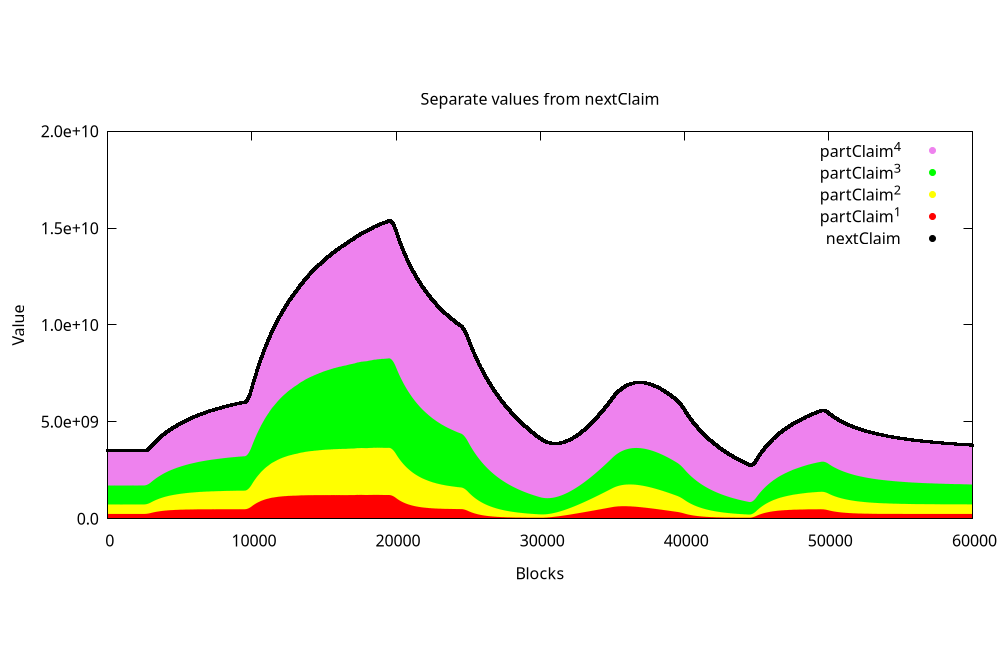}
		\caption{$\rho$ correlating with $\lambda$.}
	\end{subfigure}
	\begin{subfigure}{0.32\textwidth}
		\includegraphics[width=\textwidth]{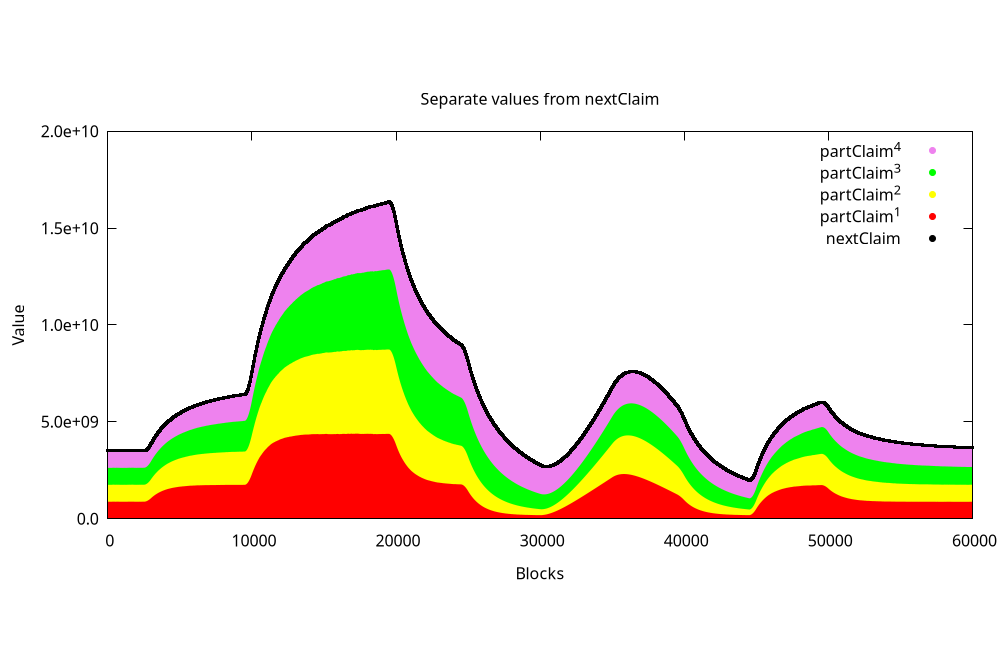}
		\caption{$\rho$ equal for every $\mathcal{FRSC}$.}
	\end{subfigure}
	\begin{subfigure}{0.32\textwidth}
		\includegraphics[width=\textwidth]{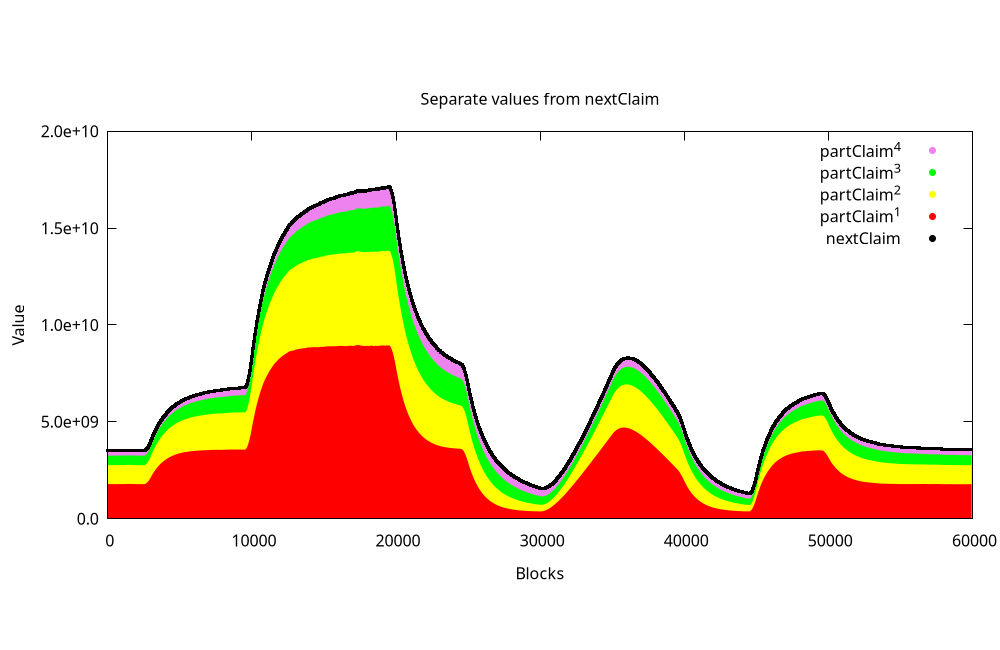}
		\caption{$\rho$ negatively correlating with $\lambda$.}
	\end{subfigure}

	\caption{Experiment II -- multiple $\mathcal{FRSC}$s using various distributions of $\rho$ and their impact on $\partial Claim$, where $\mathbb{C}$ = 0.7.}	\label{fig:percentages}
\end{figure*}

\begin{figure}
	\centering
	\begin{subfigure}{0.44\textwidth}
		\includegraphics[width=\textwidth]{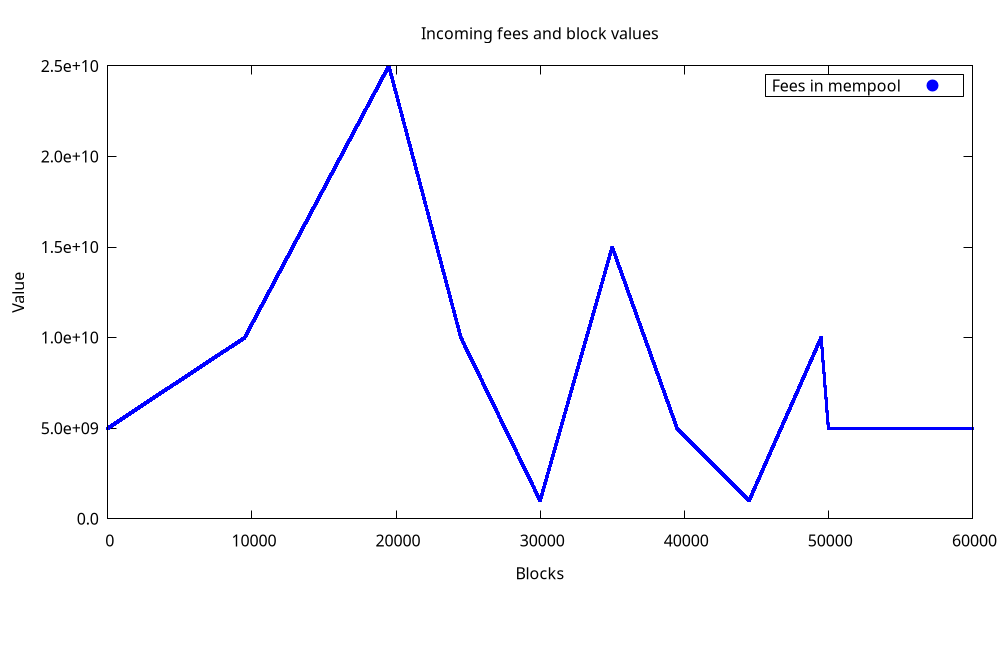}
		\caption{A custom fee scenario for Experiment III.}\label{fig:exp3-fees}
	\end{subfigure}
	\begin{subfigure}{0.44\textwidth}
		\includegraphics[width=\textwidth]{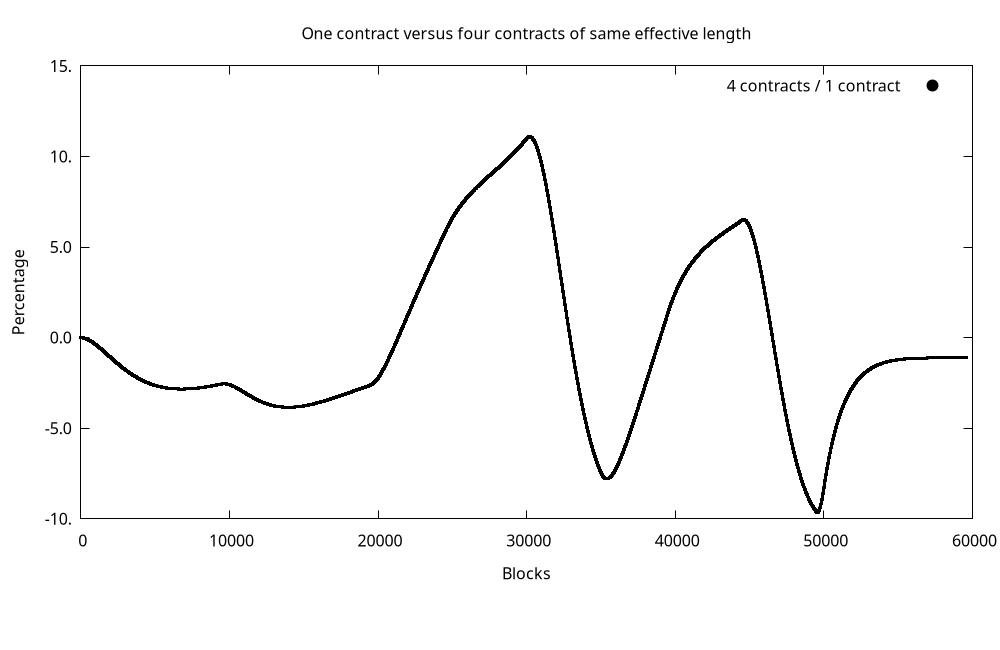}
		\caption{A relative difference in $nextClaim$ between 4 $\mathcal{FRSC}$s and a single $\mathcal{FRSC}$.}\label{fig:exp3-relative-diff}
	\end{subfigure}
	
	\caption{Experiment III comparing 4 $\mathcal{FRSC}$s and 1 $\mathcal{FRSC}$, both configurations having the same $\text{effective}\_\lambda$.  }\label{fig:effective_length}
\end{figure}

\subsection{Experiment II}
\label{sec:exp2}

\subsubsection{\textbf{Methodology}}
In this experiment, we investigated how multiple $\mathcal{FRSC}$s dealt with the same scenario as before -- i.e., varying $\mathbb{C}$.
In detail, we investigated how individual $\mathcal{FRSC}$s  contributed to the $nextClaim_{[H+1]}$ by their individual $\partial Claim^{\mathcal{FRSC}_{[H]}}_{[H+1]}$.
This time, we varied only the parameter $\mathbb{C} \in \{0.5, ~0.7, ~0.9\}$, and we considered four $\mathcal{FRSC}$s:
\begin{center}
	$\mathcal{FRSC}s=\{$\\
	$\mathcal{FRSC}^{1}(\_, 1008, 0.07), \mathcal{FRSC}^{2}(\_, 2016, 0.14),$\\
	$\mathcal{FRSC}^{3}(\_, 4032, 0.28), \mathcal{FRSC}^{4}(\_, 8064, 0.51)\},$
\end{center}
where their lengths $\lambda$ were set to consecutive powers of two (to see differences in more intensive averaging spread across longer intervals), and their redistribution ratios $\rho$ were set to maximize the potential of averaging by longer $\mathcal{FRSC}$s (see details below in \autoref{sec:different-rhos}).

\subsubsection{\textbf{Results}}
The results of this experiment are depicted in \autoref{fig:exp-II}.
We can observe that the shorter $\mathcal{FRSC}$s quickly adapted to new changes and the longer $\mathcal{FRSC}$s kept more steady income for the miner.
In this sense, we can see that $\partial Claim^{4}$ held steadily over the scenario while for example $\partial Claim^{1}$ fluctuated more.

Since the scenarios of fees evolution in the mempool was the same across all our experiments (but \autoref{sec:exp3}), we can compare the $\mathcal{FRSC}$ with $\lambda$ = 5600 from \autoref{sec:exp1} and the current setup involving four $\mathcal{FRSC}$s -- both had some similarities.
This gave us intuition for replacing multiple $\mathcal{FRSC}$s with a single one, which we further investigated in \autoref{sec:exp3}.

\subsubsection{\textbf{Different $\rho$s}}\label{sec:different-rhos}
In \autoref{fig:percentages} we investigated different values of $\rho$ in the same set of four contracts and their impact on $\partial Claim$s.
The results show that the values of $\rho$ should correlate with $\lambda$ of multiple $\mathcal{FRSC}$s to maximize the potential of averaging by longer $\mathcal{FRSC}$s.

\subsection{Experiment III}\label{sec:exp3}
\subsubsection{\textbf{Methodology}}
In this experiment, we investigated whether it is possible to use a single $\mathcal{FRSC}$ setup to replace a multiple $\mathcal{FRSC}$s while preserving the same effect on the $nextClaim$.
To quantify a difference between such cases, we introduced a new metric of $\mathcal{FRSC}s$, called $\text{effective}\_\lambda$, which can be calculated as follows:
\begin{eqnarray} \label{eq:effective_length}	
	\text{effective}\_\lambda (\mathcal{FRSC}s) = \sum_{x ~\in~ \mathcal{FRSC}s} x.\rho* x.\lambda.
\end{eqnarray}
As the example, we were interested in comparing a single $\mathcal{FRSC}$ with 4 $\mathcal{FRSC}$s, both configurations having the equal $\text{effective}\_\lambda$. 
The configurations of these two cases are as follow: 
(1) $\mathcal{FRSC}(\_, 5292, 1)$ and (2)
\begin{center}
	$\mathcal{FRSC}s=\{$\\
	$\mathcal{FRSC}^{1}(\_, 1008, 0.07), \mathcal{FRSC}^{2}(\_, 2016, 0.19),$\\
	$ \mathcal{FRSC}^{3}(\_, 4032, 0.28), \mathcal{FRSC}^{4}(\_, 8064, 0.46)\}$.
\end{center}
We can easily verify that the $\text{effective}\_\lambda$ of 4 $\mathcal{FRSC}$s is the same as in a single $\mathcal{FRSC}$ using \autoref{eq:effective_length}:
$0.07 * 1008 + 0.19 * 2016 + 0.28 * 4032 + 0.46 * 8064 = 5292$.

We conducted this experiment using a custom fee evolution scenario involving mainly linearly increasing/decreasing fees in the mempool (see \autoref{fig:exp3-fees}), and we set $\mathbb{C}$ to 0.7 for both configurations.
The new scenario of the fee evolution in mempool was chosen to contain extreme changes in fees, emphasizing possible differences.

\subsubsection{\textbf{Results}}
In \autoref{fig:exp3-relative-diff}, we show the relative difference in percentages of $nextClaim$ rewards between the settings of 4 $\mathcal{FRSC}$s versus 1 $\mathcal{FRSC}$.
It is clear that the setting of 4 $\mathcal{FRSC}$s  in contrast to a single $\mathcal{FRSC}$ provided better reward compensation in times of very low fees value in the mempool, while it provided smaller reward in the times of higher values of fees in the mempool.
Therefore, we concluded that it is not possible to replace a setup of multiple $\mathcal{FRSC}$s with a single one. 

\subsection{Experiment IV}\label{sec:exp4}
We focused on reproducing the experiment from Section 5.5 of~\cite{carlsten2016instability}. 
We were searching for the minimal ratio of \textsc{Default-Compliant} miners, at which the undercutting attack is no longer profitable strategy.
\textsc{Default-Compliant} miners are honest miners in a way that they follow the rules of the consensus protocol such as building on top of the longest chain.

We executed several simulations, each consisting of multiple games (i.e., 300k as in~\cite{carlsten2016instability}) with various fractions of \textsc{Default-Compliant} miners.
From the remaining miners we evenly created \textit{learning miners}, who learn on the previous runs of games and switch with a certain probability the best mining strategy out of the following.
\begin{compactitem}
	\item \textsc{PettyCompliant}: This miner behaves as \textsc{Default-Compliant} except one difference.
	In the case of seeing two chains, he does not mine on the oldest block but rather the most profitable block.
	Thus, this miner is not the (directly) attacking miner.

	\item \textsc{LazyFork}:
	This miner checks which out of two options is more profitable: (1) mining on the longest-chain block or (2) undercutting that block.
	In either way, he leaves half of the mempool fees for the next miners, which prevents another \textsc{LazyFork} miner to undercut him.
	
	\item \textsc{Function-Fork()}
	The behavior of the miner can parametrized with a function f(.) expressing the level of his undercutting.
	The higher the output number the less reward he receives and more he leaves to incentivize other miners to mine on top of his block.
	This miner undercuts every time he forks the chain. 
\end{compactitem}

\subsubsection{\textbf{Methodology}}
With the missing feature for difficulty re-adjustment  (present in our work as well as in~\cite{carlsten2016instability}) the higher orphan rate occurs, which might directly impact our $\mathcal{FRSC}$-based approach.
If the orphan rate is around 40\%, roughly corresponding to~\cite{carlsten2016instability}, our blocks would take on average 40\% longer to be created, increasing the block creation time (i.e., time to mine a block).
This does not affect the original simulator, as there are no $\mathcal{FRSC}s$ that would change the total reward for the miner who found the block.

Nevertheless, this is not true for $\mathcal{FRSC}$-based simulations as the initial setup of $\mathcal{FRSC}s$ is calculated with $\overline{fees} = 50$ BTC (as per the original simulations).
However, with longer block creation time and transaction fees being calculated from it, the amount of $\overline{fees}$ also changes.
Without any adjustments this results in $\mathcal{FRSC}s$ initially paying smaller reward back to the miner before they are saturated. 
To mitigate this problem, we increased the initial values of individual $\mathcal{FRSC}$s by the orphan rate from the previous game before each run.
This results in very similar conditions, which can be verified  by comparing the final value in the longest chain of our simulation versus the original simulations.
We decided to use this approach to be as close as possible to the original experiment.
This is particularly important when the parameter for the full mempool is equal to $false$ (see \autoref{sec:implementation}), which means that the incoming transaction fees to mempool are calculated based on the block creation time. 
In our simulations, we used the following parameters: 100 miners,
	10 000 blocks per game, 
	300 000 games (in each simulation run),
	exp3 learning model, and
	$\mathbb{C} = 0.7$.
Modeling of fees utilized the same parameters as in the original paper~\cite{carlsten2016instability}: the full mempool parameter disabled, a constant inflow of 5 000 000 000 Satoshi (i.e., 50 BTC) every 600s.
For more details about the learning strategies and other parameters, we refer the reader to~\cite{carlsten2016instability}.

\subsubsection*{\textbf{Setup of $\mathcal{FRSC}s$}}
Since we have a steady inflow of fees to the mempool, we do not need to average the income for the miner.
Therefore, we used only a single $\mathcal{FRSC}$ defined as $\mathcal{FRSC}$(7 056 000 000 000, 2016, 1), where the initial value of $\mathcal{FRSC}.\nu$ was adjusted according to \autoref{eq:setup}, assuming $\overline{fees} = 50$ BTC.
In next runs of any game, $\mathcal{FRSC}.\nu$ was increased by the orphan rate from the previous runs, as mentioned above.

\begin{figure}
	\centering
	\begin{subfigure}{0.40\textwidth}
		\includegraphics[width=\textwidth]{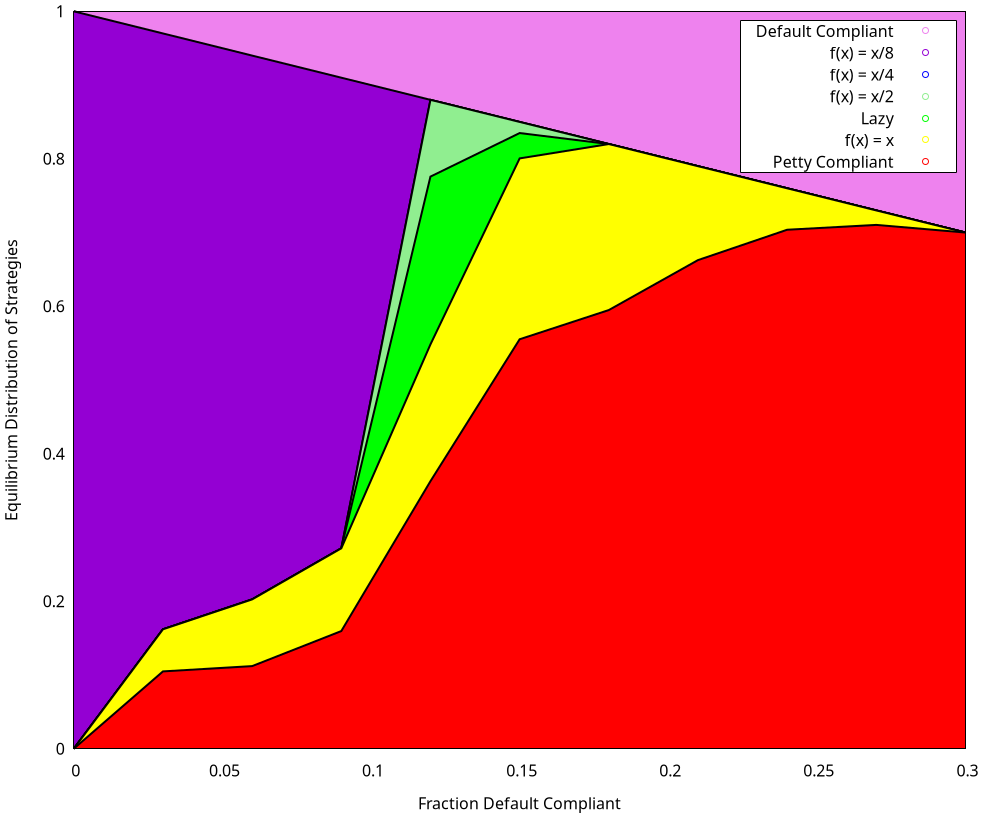}
	\end{subfigure}
	
\caption{The number of \textsc{Default-Compliant}  miners in our $\mathcal{FRSC}$ approach is $\sim$30\% (in contrast to $\sim66\%$ of~\cite{carlsten2016instability}).}\label{fig:improvement}
\end{figure}

\subsubsection{\textbf{Results}}
The results of this experiment are depicted in \autoref{fig:improvement}, and they demonstrate that with our approach using $\mathcal{FRSC}$s, we lowered down the number of \textsc{Default-Compliant} miners (i.e., purple meeting red) from the original $66\%$ to $30\%$.
This means that the profitability of undercutting miners is avoided with at least $30\%$ of \textsc{Default-Compliant} miners, indicating the more robust results.

\section{Security Analysis and Discussion}
\label{sec:discussion}
\subsection{Contract-Drying Attack}
This is a new attack that might potentially occur in our scheme; however, it is the abusive attack aimed at attacking the functionality of our scheme and not on maximizing the profits of the adversary.
In this attack, the adversary aims at getting his reward only from $\mathcal{FRSC}$s and does not include transactions in the block (or includes only a number of them).
This might result in slow drying of the funds from $\mathcal{FRSC}$s and would mean less reward for future honest miners.
Moreover, the attacker can mine in well times of higher saturation of $\mathcal{FRSC}$s and after some time decide to switch off the mining.
This might cause a deterioration in profitability for honest miners, consequently leading to deteriorated security w.r.t., undercutting attacks.

The attacker can keep repeating this behavior, which would likely lead to increased transaction fees in the mempool since the attacker keeps more un-mined transactions in the mempool than the honest miners.
Therefore, if an honest miner mines a block, he gets higher reward and at the same time deposits a higher amount from transaction fees to $\mathcal{FRSC}$s, which indicates a certain self-regulation of our approach, mitigating this kind of attack.

Additionally, we can think of lowering the impact of this attack by rewarding the miner with the full $nextClaim_{[H+1]}$ by $\mathcal{FRSC}$s only if the block contains enough transaction fees (e.g., specified by the network-regulated minimum threshold).
However, this assumes that there is always a reasonable amount of fees in the mempool, which might not be the case all the time and might result in a situation where the miners temporarily stop mining if there is not enough transactions to mine.
Nonetheless, it would require a more research to investigate this solution or a better solution mitigating this type of potential abusive attack, which we left for future work.

\subsection{Possible Improvements}
$\mathcal{FRSC}$ can contain the parameter enabling the interval of the possible change in the reward paid by $\mathcal{FRSC}^x$ (i.e., $nextClaim_{[H+1]}^{x}$) from the median of its value (computed over $\lambda$).
If $nextClaim_{[H+1]}^{x}$ would drastically increase from its median value as significantly more fees would come into the mempool, then $\mathcal{FRSC}_{[H]}^{x}$ would reward the miner with a certain value (specified by the parameter) from the interval $\langle average,nextClaim_{[H+1]}^{x} \rangle$ instead of the full $nextClaim_{[H+1]}^{x}$.
This would be particularly useful for $\mathcal{FRSC}$s with a small $\lambda$ parameter.
The parameter used for sampling might contain a stochastic function (e.g., exponential) attributing a higher likelihood of getting the values not far from the median.
However, we left the evaluation of this technique to future work.

\subsection{Adjustment of Mining Difficulty}
If the PoW blockchain with the longest chain fork-choice rule uses transaction-fee-based regime, the profitability of miners might be more volatile, which can further lead to decreased security w.r.t. undercutting attacks.
Although our solution with $\mathcal{FRSC}$s helps in mitigation of this problem, we propose another functionality that resides in adjusting the mining difficulty based on the total collected fees during the epoch.
In detail, the difficulty can be increased with higher fees collected from transactions during the epoch and vice versa.
Further research would be needed to evaluate this proposition.

\section{Related work}
\label{sec:relatedwork}
The work of Carlsten et al.~\cite{carlsten2016instability} is the inspiration for our paper, which for the first time describes undercutting attacks arising from the exponential distribution of block creation time and significant differences in transaction fees.
The authors simulated Bitcoin under transaction-fee-based regime and found that there exist the minimal threshold of \textsc{Default-Compliant} miners equal to $66\%$. 

Gong et al.~\cite{gong2022towards} argue that using all accumulated fees in the mempool regardless of the block size limit is infeasible in practice and can inflate the profitability of undercutting that was originally described in~\cite{carlsten2016instability}.
Furthermore, Houy~\cite{houy2014economics} demonstrates that a constrain on the block size limit (thus the number of transactions) has economic importance and allows transaction fees not dropping to zero.
Therefore, Gong et al.~\cite{gong2022towards} model the profitability of undercutting with the block size limit presented, which bounds the claimable fees in a mining round.
The authors presented a countermeasure that selectively assembles transactions into the new block, while claiming fewer fees to avoid undercutting.
We argue that in contrast to our approach, this solution cannot be enforced by the consensus protocol, and thus might still enable undercutting to occur.

Zhou et al.~\cite{zhou2019robust} deal with the problem of a mining gap, which is more significant when the throughput of blockchain is high. 
Therefore, the authors propose the self-adaptive algorithm to
adjust the block size every 1000 blocks and thus ensure that blocks have enough space to pack new transactions.

Even though Bitcoin-NG~\cite{eyal2016bitcoin} proposes a new consensus mechanism, it also contains an idea of splitting the transaction fees between two entities -- the current leader and the the miner of block -- which should incentivize the miner to include blocks created by the leader.
However, Bitcoin-NG is in some sense centralized and therefore, undercutting attacks are not its subject.

\section{Conclusion}
\label{sec:conclusion}
In this work, we focused on three problems related to transaction-fee-based regime of blockchains with the longest chain fork choice rule:
(1) the mining gap, (2) the possibility of undercutting attacks, and (3) the instability of mining rewards.
To mitigate these problems, we proposed the approach approximating a moving average based on the fee-redistributions smart contracts that accumulate a certain fraction of transaction fees and at the same time reward the miners from their reserves.
In this way, the miners are sufficiently rewarded even at the time of very low transaction fees, such as the beginning of the mining round, entering the mining protocol by new miners, market deviations, etc. 

Besides, our approach brings a higher tolerance to undercutting attacks, and increases the minimal threshold of \textsc{Default-Compliant} miner that strictly do not perform undercutting attack from 66\% reported in state-of-the-art to 30\%.

\IEEEtriggeratref{6}

\bibliography{ref}
\bibliographystyle{IEEEtran}

\end{document}